\documentstyle [prb,aps,epsf]{revtex}
\begin{document}
\widetext
\draft

\title{
Is the ground state of
$\alpha$-(BEDT-TTF)$_2${\sl M}Hg(SCN)$_4$[{\sl M}=K,Rb,Tl]\\ 
a charge-density wave or a spin-density wave?
}

\author{Ross H. McKenzie\cite{email}}

\address{School of Physics,
 University of New South Wales, Sydney 2052, Australia}

\date{\today}
\maketitle
\begin{abstract}
The nature of the
low-temperature phase of
the quasi-two-dimensional conductors
$\alpha$-(BEDT-TTF)$_2${\sl M}Hg(SCN)$_4$[{\sl M}=K,Rb,Tl]
is considered.
It is argued that the magnetic field dependence of
the phase diagram  is more consistent with
a charge-density wave,
rather than a spin-density wave  ground state.
The phase diagram of a charge-density wave
in a magnetic field is discussed using a
Ginzburg-Landau free energy derived from
microscopic theory.
New experiments are proposed to test the
charge-density-wave hypothesis and detect an additional
high field phase.\\
\end{abstract}



\section{Introduction}

Charge transfer salts based on the 
bis-(ethylenedithia-tetrathiafulvalene) (BEDT-TTF) 
molecule are novel quasi-two-dimensional conductors.\cite{ish,bro,wos}
 The widely studied family 
$\alpha$-(BEDT-TTF)$_2${\sl M}Hg(SCN)$_4$[{\sl M}=K,Rb,Tl]
 have a rich phase diagram depending on temperature,
 pressure, uniaxial stress, and magnetic field:
 metallic, superconducting, and density-wave phases are possible.\cite{bro}
Band structure calculations show they have co-existing
quasi-one-dimensional (open) and quasi-two-dimensional
(closed) Fermi surfaces.\cite{mori}
At ambient pressure they undergo a transition 
at a temperature $T_{DW} $  (see Table I)
 from a metal into a phase that is believed
to be a density-wave (DW).\cite{gru}
There is
currently controversy as to whether this is a spin-density wave 
(SDW) or a
charge-density wave (CDW).\cite{pratt,toyota,ath}
The purpose of this paper is to argue that the field
dependence of the phase diagram is more consistent 
with a CDW than a SDW.


Anomalies in the temperature dependence of the resistivity,\cite{sasaki4}
Hall coefficient,\cite{sasaki2} specific
heat,\cite{specific}, thermal expansion\cite{thermal}
magnetic susceptibility,\cite{sasaki5,christ2}
 and muon spin relaxation\cite{pratt} suggest
there is a phase transition at a temperature $T_{DW} $
from a metal to
a different phase.  It has been suggested that this 
is a spin-density wave because (i)
the magnetic susceptibility is anisotropic
below $T_{DW} $\cite{sasaki5,christ2}
and (ii) the muon spin relaxation changes below
$T_{DW}$.\cite{pratt} The magnitude of the relaxation 
implies a magnetic moment of
0.003 $\mu_B$.\cite{pratt}
However, unlike in other SDW systems
such as (TMTSF)$_2$PF$_6$ (which has a magnetic
moment of order $0.1\mu_B$\cite{le}), no muon spin rotation
is observed and nuclear magnetic resonance\cite{tak,kanoda,nmr2} and
electron spin resonance\cite{tsu} measurements show no line broadening or
splitting below $T_{DW}$.
The absence of these anomalies could be consistent
with the small magnetic moment.

 As a magnetic field
 is applied 
$T_{DW}$ decreases. At a field
$H_K$ (see Table 1), known as the kink field, the low
temperature phase is destroyed.
 The evidence for this is that at $H_K$  qualitative changes 
occur in angle-dependent magnetoresistance oscillations (AMRO)\cite{house}
 and magneto-oscillations in the resistance,\cite{brooks}
 magnetization and torque.\cite{christ} At low temperatures
the transition at $H_K$ becomes first-order and hysteresis is observed.



\section{CDW vs. SDW}

\begin{figure}
\centerline{\epsfxsize=9cm  \epsfbox{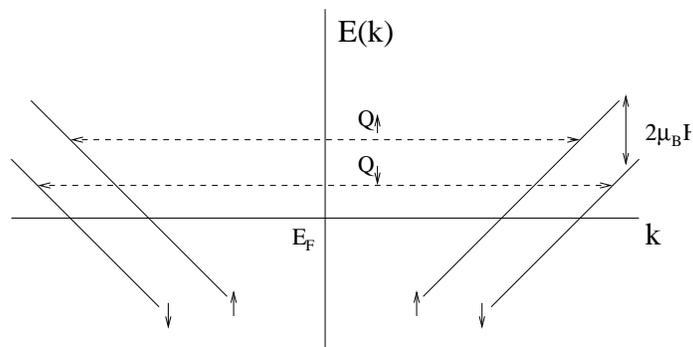}}
 \caption{
Dispersion relations of the
one-dimensional energy bands in a magnetic field.
The Zeeman splitting is $2\mu_B H.$                                        
Charge-density-wave correlations couple bands of the same spin.
Consequently, simultaneous nesting of the up-spin and
 down-spin bands is not possible
(i.e., the nesting vectors shown, $Q_\uparrow$ and $Q\downarrow$,
are unequal).
In contrast, a spin-density wave involves coupling of 
bands with opposite spin and the nesting is not
affected by a magnetic field.
\label{nesting}}
\end{figure}

If the low-temperature state is a SDW its destruction in high magnetic fields
cannot be explained in terms of mean-field theory. The SDW was proposed
to result from the nesting of the quasi-one-dimensional Fermi surface.
 However, in a SDW,
unlike a CDW, this nesting is unchanged by the Zeeman splitting of the
up and down spin
electrons (see Fig. \ref{nesting}).
 The orbital motion of the electrons in a magnetic field 
can actually improve the nesting and
thus increase the transition temperature.\cite{gor}
 To overcome this problem  that the 
SDW is destroyed at high fields it has been
proposed that magnetic breakdown\cite{osada0} or
the field enhancing quasi-one-dimensional SDW fluctuations\cite{mck2}
are responsible for the destruction of the SDW.
 However, both these effects involve the orbital
motion of the electrons and so imply that when the field is tilted at an angle
$\theta$
away from the normal to the most-conducting planes that the kink field
should vary as $H_K(\theta  = 0)/\cos \theta$.
 This is not observed:  $H_K(\theta)$ depends
very weakly on $\theta$ for
 $\theta < 50$ degrees.\cite{ath,christ,pratt2,ath2}

 \section{CDW in a magnetic field}

If the low-temperature phase is a CDW it is not difficult to explain its
destruction at high fields.  I now briefly consider the
essential physics of the theory of a CDW in a magnetic
field.\cite{Tiedje,Balseiro,Dieterich73,leung,zanchi}
For simplicity, only the case of a strictly one-dimensional system is considered
and treated at the mean-field level. 
 The only effect of a magnetic field $H$ is
to introduce a Zeeman splitting $2\mu_B H$,
where $\mu_B$ is the Bohr magneton,
 between the energies of up and down
spin electrons.  For simplicity 
the effects of fluctuations, orbital motion, and the
 quasi-two-dimensional
 Fermi surface are neglected.  The Zeeman splitting degrades
the nesting of the Fermi surface (Fig. \ref{nesting}) and reduces the mean-field 
transition temperature until it goes to zero (Fig. \ref{phased}).  

There is a simple physical argument to explain the destruction
of the CDW by a large field. It is the analogue of
an argument used to describe the Pauli or
Clogston-Chandrasekhar limit for superconductors.\cite{Shandrasekhar62}
(This is important when orbital effects can be 
neglected such as in thin films in 
a magnetic field lying in the plane of the film.\cite{fulde})
In a magnetic field the metallic state gains an
energy $-\chi H^2$ where $\chi$ is the Pauli
spin susceptibility, $\chi=\mu_B^2 \rho(E_F)$
and $\rho(E_F)$ is the density of states at the Fermi
energy. However, at zero temperature formation
of a CDW lowers the total energy of the system
by about $-\rho(E_F) \Delta(0)^2$.
Hence, at fields larger than $H \simeq 
\Delta(0)/(\sqrt{2} \mu_B) = 1.2 k_B T_{CDW}(0)/\mu_B$
the CDW will be unstable.
The results of detailed microscopic calculations
are consistent with this rough estimate (compare Fig. 2).

At low enough temperatures and high enough fields it is favourable to form
a new CDW phase, denoted CDW$_x$,
with a wavevector shifted away from $2k_F$,
 the wavevector in zero-field. 
The shift in wavevector increases with field.
 This phase is the analogue of the Fulde-Ferrell phase that is
predicted to occur in superconductors (with suppressed orbital
effects) in a high magnetic field (or strong
spin exchange)
\cite{ff,Larkin64,Dupuis93,dupuis,suz,buzdin,Maki64,Sarma63,machida}
 and the analogue of
the incommensurate phase that occurs in spin-Peierls materials in a high
field.\cite{cross,delima,bonner,kramer}
The mathematics describing these three different systems is almost
identical.

\section{Ginzburg-Landau theory}

To clarify the physics leading to the phase diagram
shown in Fig. \ref{phased} it is helpful
to formulate the description of the CDW transition
in terms of Ginzburg-Landau theory.
The simplest Ginzburg-Landau free energy functional $F[\phi]$
for a one-dimensional  system  with a complex\cite{complex} order
parameter $\phi(x)$, where $x$ is the spatial co-ordinate, is
\begin{equation}
F[\phi]=\int dx \left[
a \mid\phi\mid^2 + \ b  \mid\phi\mid^4 +
 \  c \mid {\partial \phi\over \partial x}\mid ^2 \right]
   \label{aa1}
\end{equation}
where the coefficients $a$, $b$, and $c$ can be derived from
microscopic theory
 (see the Appendix).
In order to have a stable ground state
$b$ and $c$ must be positive.
The  mean-field transition temperature
 is defined by the temperature at which
the second-order coefficient $a(T)$ vanishes.
This description is valid at low fields. $T_{CDW}$
decreases quadratically at low fields.
However, at $ \mu_B H = 1.9101 k_B T$ both the coefficients
$b$ and $c$ vanish 
signalling that new physics
becomes important.
(The fact that both $b$ and $c$ vanish
simultaneously is a consequence of
weak-coupling theory and presumably
does not occur in more general situations).

Generally, $b$ becoming negative 
denotes the transition becoming
first order\cite{tol} and $c$ becoming negative
denotes the appearance of a modulated (i.e., spatially
non-uniform) phase \cite{chaikin}.
To treat these effects the Ginzburg-Landau functional
should be expanded to higher orders in the order
parameter and its gradients.
\begin{equation}
F[\phi]=\int dx \left[
a \mid\phi\mid^2 + \ b  \mid\phi\mid^4 + e  \mid\phi\mid^6
+ \  c \mid {\partial \phi\over \partial x}\mid ^2
+ \  d \mid {\partial^2 \phi\over \partial x^2}\mid ^2
 \right]
   \label{aa2}
\end{equation}
The coefficients $d$ and $e$ must be both positive 
in order for the ground state to be stable.

We now consider only a single Fourier component,
i.e., assume $\phi(x)=\phi_x \exp(iq x)$
(this is also known as the helical state)
and minimize the free energy with respect to
both $q$ and $\phi_0$.
If $q$ is small, then $a(q)\equiv a + c q^2 + d q^4 + \cdots$
The optimum wavevector $q_x$ is then determined by
\begin{equation}
a^\prime(q_x) =0
\label{opt}
\end{equation}
and the amplitude of the order parameter is determined by
the minimum of
\begin{equation}
F[\phi_q]=
 a(q_x) \mid\phi_x\mid^2 + \ b(q_x)  \mid\phi_x\mid^4 + e(q_x) \mid\phi_x\mid^6 
\label{min}
\end{equation}

The transition between the uniform (metallic)
and the CDW$_0$ phase is determined
by $a(0)=a=0$.
Contrary, to the assumptions of some authors
this equation does {\it not}
determine the boundary between 
the CDW$_0$ and the CDW$_x$ phases.
The exact location of
that boundary is determined by finding where the free energy
of these two phases is equal.
However, $a=0$ does give its approximate location.
Since $b < 0$ for 
$ \mu_B H > 1.9101 k_B T$,
the transition from the CDW$_0$ to the CDW$_x$
phase is {\it first order,} contrary to the  claims of some.
\cite{leung,zanchi}
One also then needs to calculate the regions of
metastability of these two phases.
The boundary between the uniform (metallic) and CDW$_x$
phase is determined by $a(q_x)=0$.
I now show that this transition is {\it second order,}
contrary to what has been claimed.\cite{zanchi}
Microscopic theory implies the identity
$b(q) = {\partial^2 \over \partial q^2} a(q)$.\cite{suz}
  Hence, for small $q_x$,
 $ {\partial  \over \partial q} a    (q_x) = 2 q_x(c + 2 d q_x^2)=0$
and thus $b(q_x)= 10 d q_x^2 > 0$.

\section{Comparison with experiment}

Although there are numerous quasi-one-dimensional
materials with a CDW ground state most have transition temperatures
of the order of 100 K and so fields of the order
of several hundred tesla would be required to destroy the
CDW. However, Per$_2$[Au(mnt)$_2$]
has a transition temperature of 12 K
and was recently studied at fields up to 18 tesla.\cite{Port}
It was found that the transition temperature 
decreased quadratically with field, but at only
half the rate predicted by the mean-field theory discussed here.

\begin{figure}
\centerline{\epsfxsize=9.0cm  \epsfbox{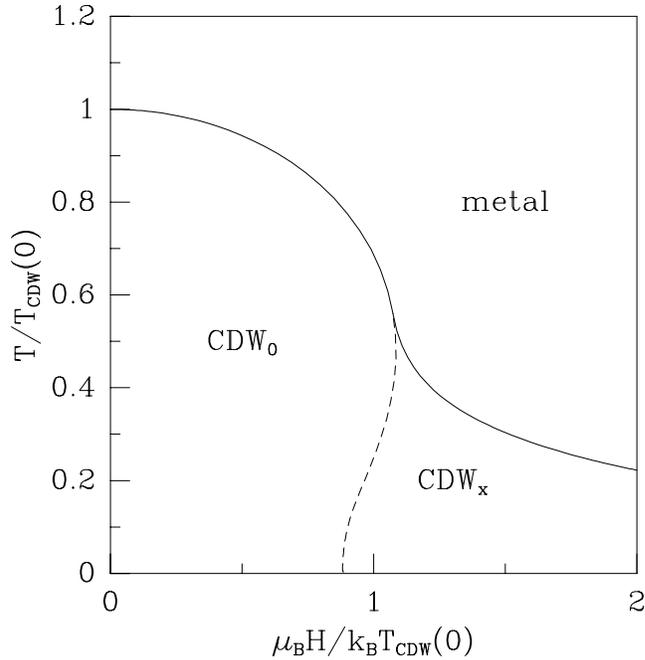}}
\vskip 0.7cm
\caption{
Phase diagram of a charge-density wave in a magnetic field. The three
phases are the metallic phase,   charge-density wave (CDW$_0$) and
 the modulated phase (CDW$_x$).  
  The solid line denotes a second-order phase transition.
  The dashed line denotes the approximate position
of a first-order phase transition.
In the CDW$_0$ phase the wavevector  $Q = 2k_F$
 and is determined by the Fermi
surface nesting.  In the CDW$_x$ phase the CDW wavevector $Q$
 shifts with
increasing field. It is an analogue of the Fulde-Ferrell state in
 superconductors and the incommensurate spin-Peierls state.
The temperature $T$ is normalized to the
mean-field transition temperature in zero-field, $T_{DW}(0)$.
\label{phased}}
\end{figure}

The phase boundary of the CDW$_0$ phase shown
in Fig. \ref{phased} is quantitatively
consistent with the phase boundary observed for the
DW phase of 
$\alpha$-(BEDT-TTF)$_2${\sl M}Hg(SCN)$_4$[{\sl M}=K,Rb,Tl].
The transition becomes first order for $T < 0.4T_{DW}(0)$.
Table I shows that the dimensionless ratio
$\mu_B H_K / k_B T_{DW}(0)$
is the same for all three salts.
It is about twice the value of 0.9 predicted by 
Fig. \ref{phased}. However, it is quite likely
that fluctuations reduce $T_{DW}(0)$ below
its mean-field value,\cite{mck} thus increasing the ratio
$\mu_B H_K / k_B T_{DW}(0)$.
If instead of of $T_{DW}(0)$
the value of the zero-temperature gap $\Delta(0)= (5 \pm 1)$
meV, estimated from magnetoresistance measurements\cite{mck1} is used
then $\mu_B H_K /\Delta(0) = 0.3$, compared to 
the theoretical value of 0.5.

\section{The modulated phase}

If the phase diagram of Fig. 2 describes
$\alpha$-(BEDT-TTF)$_2${\sl M}Hg(SCN)$_4$[{\sl M}=K,Rb,Tl] 
then one should be able to observe the modulated phase.
At low temperatures significant hysteresis is observed
near the kink field. This is consistent with
the transition between the uniform CDW and modulated CDW phases
being first order. (Although it could also be
glassy behaviour because the CDW wave vector won't line up).

There is some experimental evidence for a
phase transition at fields    above $H_K$.
First, torque measurements in the K salt show hysteresis
near 27.2 T, in addition to the hysteresis
at the kink field, of about 23 K.\cite{christ}
Secondly, the character of Shubnikov - de Haas (SdH)
oscillations changes significantly on
lowering the temperature at fields well above
$H_K$.\cite{hill}
Conventional SdH oscillations increase in
amplitude with decreasing temperature.
In contrast, in the K salt, for fields in
the range 36 to 43 tesla, the amplitude
of the oscillation at the fundamental frequency increases with decreasing
temperature and then decreases until it has a minimum
near 1 K and then increases as the temperature is 
lowered further. These results were interpreted in
terms of competition between transport involving
the bulk and the surface of the sample.
An alternative explaination is that the minimum at 1 K
is associated with a phase transition.
On the other hand, it should be pointed out this
anomaly was not seen in de Haas - van Alphen experiments.
Third, recent temperature sweeps of resistance and
torque at fixed fields in the range 25 to 28 tesla
(and thus above $H_K$) show features near 4 K.\cite{kart}
These features have been interpreted as being
due to a transition into  a new low-temperature
high-field phase. A second set of
torque and resistance measurements\cite{sasaki0}
has been used to justify an even more complicated 
phase diagram.

It should be pointed out that the 
modulated phase may have a much smaller effect
on transport properties than the
 the CDW$_0$ phase.
At low temperatures the
order parameter decreases by a factor of about three
on crossing from the  CDW$_0$ phase to the CDW$_x$ phase
(compare Fig. 7. in Ref. \onlinecite{suz}).
If there is a reconstructed quasi-two-dimensional
Fermi surface due to the CDW (this is still controversial)
then the  associated energy gap
is proportional to the CDW order parameter,\cite{mck1}
and so will also be reduced by three.
Since the magnetic breakdown field $H_0$ for the reconstructed
Fermi surface is proportional to the square of the
energy gap\cite{sho},  $H_0$ will be reduced by an
order of magnitude. In the CDW$_0$ phase $H_0 \sim 60 $ T
and so in the CDW$_x$ phase $H > 5 H_0$ and the
magnetic breakdown will be so complete
that the reconstructed Fermi surface will have little effect
on transport (compare Fig. 3 in Ref. \onlinecite{mck1}).
However, if the magnetoresistance in the
 density-wave phase is not due a reconstructed quasi-two-dimensional
Fermi surface, but due to an alternative
mechanism, such as proposed by Yoshioka\cite{yosh}
then the CDW$_x$ phase might still have an observable
effect on the magneto-transport.

Perhaps the transition into the modulated
phase from the metallic phase could be detected
by measurements of thermodynamic quantities such as specific heat
and thermal expansion.
The anomaly at low fields is very small\cite{specific}
and will be even smaller for the modulated phase.
Yet hopefully, it can still be detected on high-quality crystals.
For the spin-Peierls compound CuGeO$_3$
the transition into the modulated phase was recently detected
by thermodynamic measurements.\cite{lorenz}

\section{future directions}

In conclusion, it is argued that the magnetic field dependence of
the phase diagram of
$\alpha$-(BEDT-TTF)$_2${\sl M}Hg(SCN)$_4$[{\sl M}=K,Rb,Tl]
 is more consistent with
a charge-density wave,
rather than a spin-density wave  ground state.
Based on this work I suggest some
future directions for both theoretical 
and experimental work.
On the theory side four questions need to
be addressed.

(1) What is extent of the metastable regions of the
CDW and CDW$_x$  phases?
The calculations should be compared to 
the experimental results which show that on
down sweeps of the field hysteresis is
observed down to fields a small as half   
of the kink field.

(2)
Is the spatial dependence of the lowest energy modulated phase
of the form $\phi_0 \exp(i q_x x)$, assumed here,
or could it be $\phi_0 \cos( q_x x)$,
or the elliptic function $\phi_0  {\rm sn}( q_x x)$?
These three posibilities
are also known as the Fulde-Ferrell\cite{ff},
 Larkin-Ovchinikov,\cite{Larkin64}
and soliton states,\cite{bonner,machida} repectively
and have been considered for the analogous 
problem for superconductors and spin-Peierls systems.
This might change the order of the transition
from the CDW to the CDW$_x$ phase.\cite{machida}

(3) If the ground state 
is a CDW how does one explain the
anisotropic susceptibility and muon spin relaxation
that have been interpreted as evidence for a SDW?
It has been suggested \cite{toyota,ath2} that 
as in the purple bronze $\gamma$-Mo$_4$O$_{11}$,\cite{schlenker}
 the anisotropy
arises from Landau paramagnetism
(an orbital effect).
The muon spin relaxation data suggesting a very
small magnetic moment might be explained
by a coexisting CDW and SDW.
A possible mechanism for
such a ground state 
has been considered by Overhauser.\cite{overhauser}
It is interesting that (TMTSF)$_2$PF$_6$
was thought to be a SDW but recent
x-ray scattering measurements
suggest    a coexisting CDW and SDW.\cite{pouget}

(4) Could the ground state of
$\alpha$-(BEDT-TTF)$_2${\sl M}Hg(SCN)$_4$[{\sl M}=K,Rb,Tl] 
be a SDW and the high field transition 
be a spin flop transition?\cite{contrary}
Spin flop transitions occur in anisotropic
antiferromagnets
when the Zeeman energy becomes of the order of
the anisotropy energy. This causes a change
in the relative orientation of the spins.
Typical phase diagrams are qualitatively similar to
that shown in Fig. 2.\cite{chaikin}
For this explaination to be plausible
a theoretical model must be produced which explains 
how the reorientation of the spins will change
the electronic structure so the AMRO and magneto-oscillations
change above the kink field.\cite{kimura}

I propose several experiments which
could help resolve some of the issues raised in this paper.

(1) High resolution x-ray scattering measurements
should be done 
in order to directly observe the CDW in
$\alpha$-(BEDT-TTF)$_2${\sl M}Hg(SCN)$_4$[{\sl M}=K,Rb,Tl].

(2) In the presence of a CDW one can observe new
infrared active phonon modes, sometimes called ``phase
phonons.''\cite{ir}
The appropriate infrared measurements should be made.

(3) Specific heat and thermal expansion 
measurements should be done in high fields
in order to see whether there is a 
new phase above the kink field.

(4) A search should also be made for the
modulated phase in Per$_2$[Au(mnt)$_2$].
This will require pulsed magnetic fields of the order of
40 to 60 tesla.

\acknowledgements

This research was supported by the Australian Research Council
and the USA National High Magnetic Field Laboratory
which is supported by NSF Cooperative Agreement
No. DMR-9016241 and the state of Florida.
This work was stimulated by discussions with J.S. Brooks
who showed me unpublished results of N. Toyota.
Discussions with A. R. Bishop, P. M. Chaikin, J. Eldridge,
S. Hill, D. Huse, P. Sandhu, T. Sasaki  are gratefully acknowledged.
The figures were  produced by D. Scarratt.

\appendix

\section*{Ginzburg-Landau coefficients}

The coefficients in the Ginzburg-Landau free energy
can be calculated from microscopic theory using
standard Greens function techniques.\cite{dupuis,suz,mck}
For completeness I  give some of the expressions here  ($k_B=1$).
\begin{equation}
a(q) = { 1 \over \pi v_F} \left[ \ln 
\left( { T \over T_{MF} } \right)
+ \Psi \left( {1 \over 2} \right)
- {1 \over 2} \sum_{\alpha = \pm 1}
\Psi \left( {1 \over 2}
+  {\alpha v_F q + 2 \mu_B H \over 4 i \pi T }\right) \right]
\label{aopt}
\end{equation}
where $\Psi(z)$ is the digamma function and $v_F$ is the Fermi velocity.
\begin{equation}
b(0) = { -1 \over 4\pi v_F} 
{1 \over (2 \pi T)^2 }
\Psi^{\prime \prime}  \left( {1 \over 2}
+  {\mu_B H \over 2 i \pi T }\right)
  \label{bopt}
 \end{equation}




\begin{table}
\caption{
Values for the zero-field transition temperature
$T_{DW}(0)$ and the kink field $H_K$ for
the different 
$\alpha$-(BEDT-TTF)$_2${\sl M}Hg(SCN)$_4$[{\sl M}=K,Rb,Tl] 
salts.  The dimensionless ratio $\mu_B H_K / k_B
T_{DW}(0)$ can be compared to the
value of 0.9 corresponding to the low-temperature
boundary
of the CDW$_0$ phase in the
 phase diagram shown in Fig. \protect\ref{phased}.
The actual ratio is expected to be smaller than
0.9 because the actual transition temperature
is reduced below the mean-field value by fluctuations \protect\cite{mck,mck1}.
}
\begin{tabular}{cccc}
M& $T_{DW}(0)$ (K) & $H_K$ (T) & $\mu_B H_K / k_B T_{DW}(0)$ \\ 
\tableline
K & 8 & 23 & 1.9 \\
Tl & 9 & 27 & 2.0 \\
Rb & 12 & 32 & 1.8 \\
\end{tabular}
\label{table1}
\end{table}


\begin{references}

\bibitem[*]{email}electronic address: ross@newt.phys.unsw.edu.au

\bibitem{ish} For a review, T. Ishiguro and K. Yamaji,
{\it Organic Superconductors}, Second edition
(Springer-Verlag, Berlin, 1997).

\bibitem{bro} J. S. Brooks,
Mat. Res. Soc. Bull. {\bf 18}, 29 (1993).

\bibitem{wos} 
J. Wosnitza,
{\it Fermi Surfaces of Low Dimensional
 Organic Metals and Superconductors}
(Springer Verlag, Berlin, 1996).

\bibitem{mori} H. Mori {\it et al.},
Bull. Chem. Soc. Jpn. {\bf 63}, 2183 (1990);
L. Ducasse and A. Frisch,
Solid State Comm. {\bf 91}, 201 (1994);
R. Rousseau {\it et al.}, 
J. Phys. (France) I. {\bf 6},    1527 (1996).

\bibitem{gru} For an introduction see
 G. Gr\"uner, {\it Density Waves in Solids},
(Addison-Wesley, Redwood City, 1994).

\bibitem{pratt}
F. L. Pratt {\it et al.},
 Phys. Rev. Lett. {\bf 74}, 3892 (1995).

\bibitem{toyota} J. S. Brooks {\it et al.}, in
{\it Physical Phenomena at High Magnetic Fields II},
edited by Z. Fisk {\it et al.},
(World Scientific, Singapore, 1996).

\bibitem{ath} G.J. Athas, Ph.D thesis, Boston University, 1996
(unpublished).

\bibitem{sasaki4} T. Sasaki {\it et al.},
Solid State Comm. {\bf 75}, 93  (1990).

\bibitem{sasaki2} T. Sasaki, S. Endo,  and N. Toyota,
Phys. Rev. B {\bf 48}, 1928 (1993).

\bibitem{specific}P. F. Henning {\it et al.},
Solid. State  Commun. {\bf 95}, 691 (1995);
A. E. Kovalev and H. M\"uller,
Synth. Met.       {\bf 86}, 1997 (1997).

\bibitem{thermal}
M. K\"oppen {\it et al.},
Synth. Met. {\bf 86}, 2059 (1997).

\bibitem{sasaki5} T. Sasaki, H. Sato, and N. Toyota,
Synth. Met.       {\bf 41-43}, 2211 (1991).

\bibitem{christ2} P. Christ { \it et al.},
Synth. Met. {\bf 86}, 2057 (1997).

\bibitem{le}
L. P. Le {\it et al.},
Phys. Rev. B {\bf 48}, 7284 (1993).

\bibitem{tak} T. Takahashi {\it et al}.,
 Synth. Met. {\bf 55-57}, 2513   (1993).

\bibitem{kanoda} K. Kanoda {\it et al}.,
 Synth. Met. {\bf 70}, 973   (1995).


\bibitem{nmr2} K. Miyagawa, A. Kawamoto, and K. Kanoda,
Phys. Rev. B {\bf 56}, R8487 (1997).

\bibitem{tsu} R. Tsuchiya {\it et al}.,
 Synth. Met. {\bf 70}, 965   (1995).

\bibitem{house}
A. A. House { \it et al.},
 J. Phys.: Cond. Matter {\bf 8},  8829 (1996).

\bibitem{brooks} J. S. Brooks {\it et al.},
Phys. Rev. Lett. {\bf 69}, 156  (1992);
 Phys. Rev. B {\bf 52}, 14457 (1995).


\bibitem{christ} P. Christ { \it et al.},
 Physica    B {\bf 204},  153 (1995).


\bibitem{gor}L. P. Gor'kov and A. G. Lebed,
 J. Phys. Lett. (Paris) {\bf 45}, L433 (1984).

\bibitem{osada0} T. Osada, S. Kagoshima, and N. Miura,
Synth. Met.   {\bf 70}, 931 (1995).

\bibitem{mck2}
R. H. McKenzie, Phys. Rev. Lett.
{\bf 74}, 5140  (1995).

\bibitem{pratt2}
F. L. Pratt {\it et al.},
 Phys. Rev. B     {\bf 45}, 13904 (1992).

\bibitem{ath2} G.J. Athas {\it et al}.,
 Synth. Met. {\bf 70}, 843   (1995).

\bibitem{Tiedje}T. Tiedje, J.F. Carolan and
A. J. Berlinsky, Can. J. Phys. {\bf 53}, 1593 (1975).

\bibitem{Balseiro}C. A. Balseiro and L. M. Falicov, Phys. Rev. B 
{\bf 34}, 863 (1985).

\bibitem{Dieterich73} W. Dieterich and P. Fulde, Z. Phys. {\bf 265}, 239
(1973).
         
\bibitem{leung} M. C. Leung,
 Phys. Rev. B {\bf 11}, 4272 (1975).

\bibitem{zanchi}   
D. Zanchi, A. Bjelis, and G. Montambaux, Phys. Rev. B {\bf 53}, 1240 (1996).
                          
\bibitem{Shandrasekhar62} B.S. Shandrasekhar, Appl. Phys. Lett. {\bf 1}, 7
(1962); A.M. Clogston, Phys. Rev. Lett. {\bf 9}, 266 (1962).

\bibitem{fulde} P. Fulde, Adv. Phys. {\bf 22}, 667 (1973).

\bibitem{ff}
 P. Fulde and R.A. Ferrell, Phys. Rev. {\bf 135}, A550 (1964).

\bibitem{Larkin64} A.I. Larkin and Yu.N. Ovchinnikov, Sov. Phys. JETP
{\bf 20}, 762 (1965).

\bibitem{Dupuis93}
N. Dupuis and G. Montambaux, Phys. Rev. B {\bf 49}, 8993 (1994).
                          
\bibitem{dupuis}
N. Dupuis, Phys. Rev. B {\bf 51}, 9074 (1995).
                          
\bibitem{suz} Y. Suzumura and K. Ishino,
 Prog. Theor. Phys. {\bf 70}, 654 (1983).

\bibitem{buzdin}
A. I. Buzdin and H. Kachkachi, Phys. Lett. A {\bf 225}, 341 (1997).

\bibitem{Maki64} K. Maki and T. Tsuneto, Prog.    Theor. Phys. 
{\bf 31}, 945 (1964).

\bibitem{Sarma63} G. Sarma, J. Phys. Chem. Solids {\bf 24}, 1029
(1963).

\bibitem{machida}
K. Machida and H. Nakanishi,
 Phys. Rev. B {\bf 30 }, 122  (1984).

\bibitem{cross} M. C. Cross,
 Phys. Rev. B {\bf 20 }, 4606 (1979).

\bibitem{delima} R. A. T. de Lima and C. Tsallis,
 Phys. Rev. B {\bf 27 }, 6896 (1983).

\bibitem{bonner} J. C. Bonner {\it et al.},
 Phys. Rev. B {\bf 35}, 1791 (1987).

\bibitem{kramer} V. Kiryukhin {\it et al.},
 Phys. Rev. Lett. {\bf 76}, 4608 (1996); Phys. Rev. B {\bf 54 }, 7269 (1996).



\bibitem{complex} If the Fermi wavevector $k_F$
for the quasi-one-dimensional Fermi surface is
commensurate with the lattice then the CDW order
parameter will be real. This will produce some
change in the analysis that follows (e.g. the
modulated phase cannot involve the helical state)
but the overall physical picture including the phase
diagram will be similar.  
            
\bibitem{tol} J. C. Toledano and P. Toledano, {\it The Landau theory
of phase transitions: application to structural, incommensurate,
magnetic, and liquid crystal systems}, (World Scientific, Singapore, 1987),
 p.  167.

\bibitem{chaikin}
P. M. Chaikin and T. C. Lubensky,
{\it Principles of Condensed Matter Physics},
(Cambridge, Cambridge, 1995).

\bibitem{Port}G. Bonfait {\it et al.},
 Physica B {\bf 211}, 297 (1995);
M. J. Matos {\it et al.},
 Phys. Rev. B {\bf 54}, 15307 (1996).

\bibitem{mck}
R. H. McKenzie,
 Phys. Rev. B {\bf 52}, 16428 (1995).

\bibitem{mck1}
R. H. McKenzie {\it et al.}, Phys. Rev. B
{\bf 54}, R8289 (1996).

\bibitem{hill}
S. Hill {\it et al.}, Phys. Rev. B
{\bf 55}, R4891 (1997).

\bibitem{kart}
M. V. Kartsovnik {\it et al.},
Synth. Met. {\bf 86}, 1933 (1997).


\bibitem{sasaki0}
T. Sasaki        {\it et al.},
Phys. Rev. B {\bf 54}, 12969 (1996).
The high temperature phase boundary is identified
with the minima in the resistance verse temperature
curve at different fields and with the temperature
at which the torque reaches an arbitrarily defined
threshold value.

\bibitem{sho}
D. Shoenberg, {\it Magnetic Oscillations in
 Metals}, (Cambridge, Cambridge, 1984).

\bibitem{yosh}
D. Yoshioka,
J. Phys.   Soc.  Jap. {\bf 64}, 3168 (1995).

\bibitem{lorenz}
T. Lorenz {\it et al.}, Phys. Rev. B
{\bf 55}, 5914  (1997).

\bibitem{schlenker}
C. Schlenker,
{\it Low dimensional electronic properties
of molybdenum bronzes and oxides,}
(Kluwer, New York, 1989).

\bibitem{overhauser}
A. W. Overhauser,
Phys. Rev. B     {\bf 29}, 7023 (1990).

\bibitem{pouget} J. P. Pouget,
J. Phys. (France) I. {\bf 6}, 1501 (1996);
J. P. Pouget and S. Ravy, 
Synth. Met. {\bf 85}, 1523 (1997).

\bibitem{contrary}
An argument against the spin flop idea
is that such a transition occurs in
the (TMTSF)$_2$X salts at field of less
than one tesla and seems to have little
effect on the electronic properties.
\protect\cite{gru}

\bibitem{kimura}
Insight might be gained from
considering how spin orientation affects the
interlayer magnetoresistance of layered 
manganese perovskites
[T. Kimura      {\it et al.}, Science {\bf 274}, 1698 (1996)].

\bibitem{ir} C. C. Homes and J. E. Eldridge,
Phys. Rev. B {\bf 42}, 9522 (1990), and
references therein.

%

















\end{references}
\end{document}